# One-dimensional spin-½ Heisenberg antiferromagnet in a weak external magnetic field

Ping Sun and D. Schmeltzer

*Department of Physics, The City College of CUNY, Convent Avenue at 138th Street, New York, New York 10031*

(Received 9 August 1999)

The one-dimensional spin-1/2 Heisenberg antiferromagnet in a weak magnetic field is studied using the bosonization method. We derive a set of renormalization-group equations. The fixed point is reached when the field is scaled to the value at which the system is quarter filled. As the magnetic field varies, a continuum line of fixed points is formed. We compute the uniform longitudinal susceptibility $\chi_z(h)$. The singular behavior of the spin-spin correlations $\chi_z(h)$ as $h \to 0$ is found to be contained in $1/\ln(h_0/h)$ with $h_0$ a nonuniversal constant. The spin-spin correlations in the magnetic field are calculated.

Advances in technology give the studies of the one-dimensional (1D) antiferromagnetic (AF) quantum spin systems a new motivation, besides the persistent interest from theoretical and numerical viewpoints. Synthetic materials containing highly anisotropic electronic exchange interactions are made in the laboratories and exhibit interesting magnetic properties. The latest among them is the spin ladder system[1] whose low-energy excitations are gapless or gapped from the ground state depending on the number of legs being odd or even, respectively. This kind of behavior can be traced back to the difference between AF Heisenberg chains with half odd-integer spins and integer spins[2]. By applying an external magnetic field, magnetization plateaus have been observed experimentally[3,4] and numerically[5-10] and explained based on the symmetry of the systems.[11] Field transitions between gapped and gapless phases have been observed. It has been found that after the spin gap is overcome at a critical field $H = H_c$, the magnetization grows according to $M \propto \sqrt{H - H_c}$.[17]

Both the systems, the spin-½ AF chain and the spin-½ ladder with $N$ legs, can be described by the bosonization method.[19-22] The method is based on the observation[23] that for a single spin-1/2 AF Heisenberg chain the low-energy excitations are those spin-density waves (SDW) carrying momenta around "0" and "$\pi/a$" where "$a$" is the lattice spacing. These SDW's are the bosons in the effective-field theory description and the interactions between them are of the sine-Gordon form. A spin chain (ladder) with $S > 1/2$ ($N > 1$) can be described by coupling in a proper way $2S(N)$ such bosons. In order to understand how the systems respond to weak magnetic field, we need first to investigate how a single chain may behave. In zero external field we know quite a lot about the system. Due to the Bethe ansatz[24] we know the energy spectrum and the susceptibility.[25,26] Due to the bozonization and the interactions, the spin-spin correlations behave as $\langle \vec{S}(R) \cdot \vec{S}(0) \rangle \sim (-1)^R \sqrt{\ln R}/R$. The study has been extended to finite temperatures for both the Bethe ansatz[29] and the field theory.[1,14] On the other hand, our knowledge of the system in a finite magnetic field is quite limited. The magnetic properties have been investigated theoretically by the Bethe ansatz.[25] the inverse scattering method,[30] and the finite-size scaling analysis.[31] The results were compared with the experiments.[32] But the singular behavior of the susceptibility $\chi_z(H)$ as the magnetic field $H$ approaches zero has not been explained. We also do not know how the spin-spin correlations in a finite magnetic field.

In this paper we are going to answer the questions raised above. We will first discuss briefly the bozonization of the chain in an external magnetic field. The procedure involves first mapping the spin model to a spinless fermion system and then bozonizing the fermions. Applying a finite magnetic field results in changing the magnetization and so, the number of the fermions. The bozonized Hamiltonian is then modified by the field. Instead of a standard sine-Gordon model, a biased term, which to the lowest order is linear in the field, appears inside the cosine potential. The spin-wave speed $v$ and the boson stiffness $K$ are also field dependent. But since these dependences do not depend on the direction of the field, these quantities are found to be to second order when the field is weak. We then present the renormalization-group (RG) flow equations which are derived by perturbation up to second order of the Umklapp interaction. It is found that up to linear order of the field, the RG flow is still on the critical line at first. Then at a length scale determined by the magnetic field $1/h$, the system crosses over to a second region where the cosine potential scales to zero due to the fast oscillation induced by the magnetic field. $h$ is the magnetic field measured in units of the exchange energy $J$, $h = g\mu_B|H|/J$. The fixed point is reached when the magnetic field is scaled to the value at which the system is at quarter filling $k_F = 3\pi/(4a)$. The fixed point moves on a continuum line of free boson models when the initial value of the magnetic field is varied. We find that the uniform susceptibility has an infinite slope as the magnetic field goes to zero, $\chi(h) \propto 1 + 1/[2\ln(h_0/h)]$ ($h_0$ is a positive nonuniversal constant). In the infrared limit where the length scale $R > h_0/h$ we calculate the spin-spin correlations using the free boson Hamiltonian at the fixed point. In the opposite limit, $R < h_0/h$, in order to show the difference between the transversal and longitudinal correlations, we take into account the deviation of the RG flow from the critical line to the massless phase, while the crossover is the same as before. This is done by including the second-order field corrections in the RG flow. We calculate the correlations using the Callan-Symmanzik equation and in the limit the field goes to zero recover the known results. The effect of a finite temperature on the system is discussed at the end of the paper. We suggest to use the measurement of $\chi(h)$ to determine the exchange interaction $J$ of 1D Heisenberg antiferromagnets.

We start bosonization by using the Jordan-Wigner transformation which maps the spins to fermions.[33] The Heisenberg model $H = J\Sigma_j \vec{S}_j \cdot \vec{S}_{j+1} + g\mu_B H \Sigma_j S_{j,z}$ becomes

$$H = -\frac{J}{2} \sum_j (C_{j+1}^\dagger C_j + C_j^\dagger C_{j+1}) + J\sum_j \left(n_j - \frac{1}{2}\right)$$
$$\times \left(n_{j+1} - \frac{1}{2}\right) + g\mu_B H \sum_j \left(n_j - \frac{1}{2}\right). \quad (1)$$

The anticommuting operators $C_j$ are for the pseudofermions and $n_j$ the number operators, $n_j = C_j^\dagger C_j$. In the absence of the field $H$, the free part of the above Hamiltonian describes a fermion system at half filling $k_F^0 = \pi/(2a)$. From the spin viewpoint, this means that the ground-state magnetization per site $M_z = 0$. In the presence of the density-density interaction, the picture of Fermi liquid is not quite applicable in 1D and should be replaced by that of a Tomonaga-Luttinger liquid.[34] But the quantity $k_F$ still has its significance not only as the starting point of bosonization, but also as a measurable quantity. It reflects the spatial periodicity of the staggered SDW's which enters the dynamical susceptibility as the locations of the $\pm 2k_F$ peaks when the frequency is zero.[20,13] It is also related to the change of particle number, and hence the magnetization, by

$$M_z(H) = \frac{g\mu_B}{\pi}[k_F(H) - k_F^0]. \quad (2)$$

A finite field $H$ shifts the fermi surface away from half filling and the magnetization is then finite. The vacuum of the system is accordingly changed such that the normal order of the operators $n_j$ is now defined as $:n_j := n_j - k_F(H)/\pi$. With the above changes, we can use the standard operator transformation to bosonize the fermionic fields: $C_j \rightarrow \psi(x) \propto e^{ik_F x}\psi_+(x) + e^{-ik_F x}\psi_-(x)$, with $\psi_+(x) = (1/\sqrt{2\pi a})e^{i\sqrt{4\pi}\phi_+(x)}$ and $\psi_-(x) = (1/\sqrt{2\pi a})e^{-i\sqrt{4\pi}\phi_-(x)}$. In the expressions we have taken the continuum limit and represented the coordinate of the $j$th site to be "$x$." The operators $\phi_\pm$ are the chiral bosonic fields. For details of the procedure see Ref. 19. The bosonized Hamiltonian is

$$H = \frac{v}{2} \int dx \left[ K_0 (\partial_x \theta)^2 + \frac{1}{K_0}(\partial_x \phi)^2 \right]$$
$$+ g_0 \int dx \cos[qx + \sqrt{16\pi}\phi(x)] \quad (3)$$

with $g_0 = 2Ja/(2\pi\alpha)^2$ as the coupling constant, $v = Ja$ $(1+1/\pi)\sqrt{1-d^2}$, $K_0 = \sqrt{(1-d)/(1+d)}$, and $d = (2/1+\pi)\sin(k_F a)$. The momentum $q$ in the Umklapp term represents the deviation of the fermi surface, $q = 4k_F - 4k_F^0$. The SDW speed $v$ and the boson stiffness $K_0$ are in agreement with those given in Ref. 33 when $H = 0$. In a weak magnetic field their dependences on $H$ are found to be second order. When the field $H = 0$, and so $q = 0$, the Hamiltonian Eq. (3) has been studied using the RG method.[35] The initial values of the scaling variables, $x_0 = 4K_0 - 2$ and $g_0$, were found to reside on the critical line and scale to the fixed point, $(x^*, g^*) = (0,0)$. The initial value of $x$ is shifted by the field such that

$$x_0^2(h) - x_0^2 \simeq \left(\frac{h}{h_1}\right)^2, \quad (4)$$

where $h_1$ is a positive constant. The Hamiltonian Eq. (3) is defined at the microscopic length scale, the lattice spacing "$a$." The field $\theta$ is dual to the bosonic field $\phi$. These fields obey periodic boundary conditions $\theta(x+L) = \theta(x), \phi(x+L) = \phi(x)$, where the length of the system $L \rightarrow \infty$. The coupling constant $g_0$ is defined on the lattice with momentum cutoff $\Lambda = 2\pi/a$. When the cutoff $\Lambda$ is reduced to $\Lambda/b(b>1)$ and an integration in the momentum shell $[\Lambda/b, \Lambda]$ is performed one generates new terms of the form $\{\partial_x [\phi(x) + qx]\}^{2n}(n \geq 2)$ and $\cos\{2m(\sqrt{16\pi}\phi(x) + qx]\}(m \geq 1)$. As a result we find that the renormalized Hamiltonian at the scale $b > 1$ takes the form

$$H(b) = \frac{v(b)}{2} \int dx \left[ K(b)(\partial_x \theta)^2 + \frac{1}{K(b)}(\partial_x \phi)^2 \right]$$
$$+ g_0(b) \int dx \cos[(qx)z(b) + \sqrt{16\pi}\phi(x)]$$
$$+ \sum_{n=1}^\infty (\lambda_{2(n+1)}(b)(\partial_x \phi)^{2(n+1)}$$
$$+ g_{2n}(b)\cos\{2n[(qx)z(n) + \sqrt{16\pi}\phi(x)]\})$$
$$+ \sum_{m,n=1}^\infty \tilde{\lambda}_{2m,2n}(b)(\partial_x \phi)^{2n} \cos\{2m[(qx)z(b)$$
$$+ \sqrt{16\pi}\phi(x)]\}. \quad (5)$$

The Hamiltonian in Eq. (5) is characterized by: $z(b=1) = 1$ and $\lambda_{2(n+1)}(b=1) = g_{2n}(b=1) = \tilde{\lambda}_{2m,2n}(b=1) = 0$. We introduce the new variables $x(b) = 4K(b) - 2$ and $y(b) \propto g_0(b)$ and find the following RG equations for the coupling constants:

$$\frac{dx}{d\ln b} = -y^2, \quad \frac{dy}{d\ln b} = -xy, \quad \frac{dz}{d\ln b} = -cy^2, \quad (6)$$

where $c = \sqrt{2/\pi}$. The proportionality constant in defining $y(b)$, which is nonuniversal, is chosen in such a way that the coefficient on the right side of the first equation is of magnitude "1." The rest of the parameters have negative dimensions and are therefore irrelevant, $d \ln \lambda_{2(n+1)}/d \ln b < 0$, $d \ln g_{2n}/d \ln b < 0$, and $d \ln \tilde{\lambda}_{2m,2n}/d \ln b < 0$. In addition to Eqs. (6) we have two more relations: $v_0 K_0 = v(b)K(b)$ and the scaling equation for the momentum $q$ ($\propto H$ to the linear order in the magnetic field), $q(b) = bqz(b)$. We investigate the flow in the gapless phase $x^2(b) - y^2(b) \geq 0$ for weak magnetic fields $h < 1$. From the scaling Eqs. (6) we observe that $q(b)$ grows with the scaling. At the scale $b = b_0$ such that

$$q(b_0) = b_0 qz(b_0) = \pi/a, \quad (7)$$

the Umklapp term becomes irrelevant. This corresponds to $q(b_0) = 4[k_F(H) - \pi/(2a)] = \pi/a$, or $k_F(H) = 3\pi/(4a)$. At this point, the Umklapp interaction is $J\Sigma_j(-1)^j\cos[\sqrt{16\pi}\phi(x_j)] \simeq g(b_0)\int dx \sin[\sqrt{16\pi}\phi(x)]\partial_x\phi$. For scales $b \geq b_0$ we obtain the following model ($s \geq 1$):

$$
\begin{aligned}
H(s) = \frac{\bar{v}}{2} \int dx & \left[ \bar{K}(s)(\partial_x\theta)^2 + \frac{1}{\bar{K}(s)}(\partial_x\phi)^2 \right] \\
& + G(s) \int dx \sin[\sqrt{16\pi}\phi(x)]\partial_x\phi \\
& + G_2(s) \int dx \cos[2\sqrt{16\pi}\phi(x)] \\
& + \bar{\lambda}_{2,2}(s) \int dx \cos[2\sqrt{16\pi}\phi(x)](\partial_x\phi)^2 + \dots,
\end{aligned}
$$
(8)

where $G(s=1) = g(b_0)\sqrt{16\pi}$, $\bar{K}(s=1) = K(b_0)$, $G_2(s=1) = g_2(b_0)$, and $\bar{\lambda}_{2,2}(s=1) = \bar{\lambda}_{2,2}(b_0)$. From Eq. (8) we observe that the new coupling constants $G$, $G_2$, and $\bar{\lambda}_{2,2}$ are strongly irrelevant, which result in the absence of scaling of $\bar{K}(s)$, $d\bar{K}(s)/d\ln s = 0$. Hence we identify a continuum set of fixed points $\bar{K}(s)_{s \to \infty} \to K(b_0)$, controlled by the value of $b_0$ and the initial value of the magnetic field $h$. As a result we obtain the longitudinal magnetic susceptibility $\chi_z(h) = 2(g\mu_B)^2K(b_0)/(J\pi^2)$ where $K(b_0) = \frac{1}{2}[1 + \frac{1}{2}x(b_0)]$, or

$$
\chi_z(h) = \frac{(g\mu_B)^2}{J\pi^2}\left[1 + \frac{1}{2}x(b_0)\right].
$$
(9)

In order to clarify the singular behavior in Eq. (9) we restrict ourselves to the critical line $x(b=1) = x_0$ and $y(b=1) = y_0$, such that $x^2 - y^2 = 0$. This is valid by neglecting the second-order field dependence in Eq. (4). Integrating Eqs. (6), we obtain

$$
x(b_0) = \frac{x_0}{1 + x_0 \ln(b_0)}.
$$
(10)

From the conditions $q(b_0) = \pi/a$ and $q \propto h$ we find

$$
\ln(b_0)\left[1 - \frac{cx_0^2}{1 + x_0 \ln(b_0)}\right] \simeq \ln\left(\frac{1}{h}\right).
$$
(11)

Equation (10) can be solved in two critical regions: (i) $h \to 0$ such that $x_0\ln(b_0) > 1$. From Eq. (11) we obtain $\ln b_0 \simeq cx_0 + \ln(1/h) \equiv \ln(h_0/h)$. Hence $x(b_0) \simeq 1/\ln b_0 \simeq 1/\ln(h_0/h)$. (ii) The nonuniversal region, $x_0\ln(b_0) < 1$. From Eq. (11) we obtain $\ln b_0 \simeq (1 - cx_0^2)^{-1}\ln(1/h)$, giving rise to $x(b_0) \simeq x_0\{1 + [x_0/(1 - cx_0^2)]\ln h\} \simeq x_0 h^{\nu_0}$ where $\nu_0 = x_0/(1 - cx_0^2) > 0$ is a nonuniversal exponent. We introduce the crossover magnetic field $\bar{h} = \exp(-1/x_0)$ (which depends on $x_0$), and obtain the susceptibility

$$
\chi_z(h) = \frac{(g\mu_B)^2}{J\pi^2}\begin{cases} 1 + \dfrac{1}{2}\dfrac{1}{\ln(h_0/h)} & h < \bar{h} \\[2mm] 1 + \dfrac{1}{2}x_0 h^{\nu_0} & h > \bar{h}, \end{cases}
$$
(12)

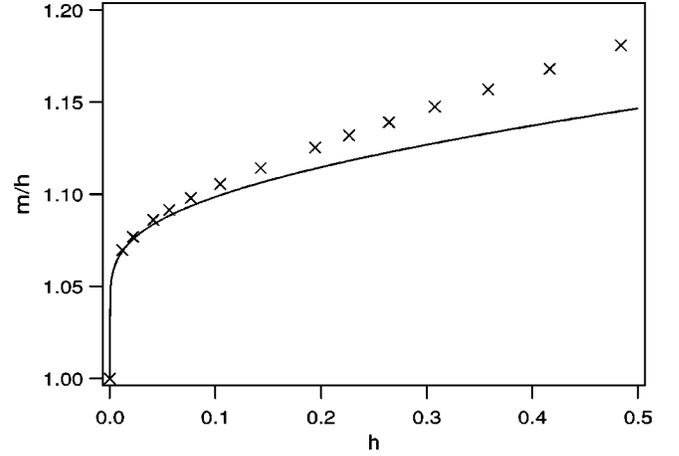

FIG. 1. The plot of $m(h)/h$ vs $h$. The solid line is drawn using Eq. (13) at $h_0 = 6.8$. The crosses are the numerical data taken from Ref. 25.

where $h_0 \equiv \exp(cx_0)$. We can now compare our susceptibility with the numerical data given in Ref. 25. We consider the limit $h \to 0$ where the lowest-order logarithm contribution is the major correction. Integrating the above equation in the region $h < \bar{h}$, we obtain the dimensionless magnetization

$$
m(h) = \frac{J\pi^2}{(g\mu_B)^2}\int_0^h dh\chi_z(h) = h + \frac{h_0}{2}E_1\left[\ln\left(\frac{h_0}{h}\right)\right],
$$
(13)

where $E_1(x)$ is the exponential-integral function defined as $E_1(x) = \int_1^\infty e^{-xt}dt/t$ for $x > 0$. We find the nonuniversal constant $h_0 = 6.8$ (see Fig. 1). From the figure we see that the expression gives the asymptotic behavior of $m/h$ as the field $h$ goes to zero. To get more precise value of the constant $h_0$, magnetization data at fields closer to zero are needed.

We are now in the position to calculate the spin-spin correlations in the magnetic field. We concentrate on two limits. The first is the infrared limit where the length scale $R > b_0 \sim h_0/h$. In this limit the problem is controlled by the free boson spin liquid with the boson stiffness given by the fixed point value $K(b_0)$. As a result the transversal spin correlation is determined by the correlation of the boson field $\theta$ with the exponent $1/[2K(b_0)]$. The longitudinal part is controlled by the bosonic field $\phi$ with the exponent $2K(b_0)$. So we have

$$
\langle S_+(R)S_-(0)\rangle_{R>b_0} = \frac{\cos[2k_F(h)aR]}{R^{1-1/[2\ln(h_0/h)]}}\text{ const,}
$$

$$
\langle[S_z(R) - M_z][S_z(0) - M_z]\rangle_{R>b_0} = \frac{\cos[2k_F(h)aR]}{R^{1+1/[2\ln(h_0/h)]}}\text{ const,}
$$
(14)

where the Fermi momentum $k_F(h)$ is related to the magnetization by Eq. (2). The second is the nonuniversal limit $R < b_0$. This region is in principle not accessible by the RG since the irrelevant coupling constants, which are not involved in the RG Eqs. (6), contribute. We will treat them perturbatively by keeping only the most important contributions, those from the Umklapp interaction in Eq. (3), in the following calculation. We also make use of the fact that un-der this limit, the magnetic field can be ignored in the Umklapp term in Eq. (3). The only effect of the magnetic field is given by the shift of $x_0$ to $x_0(h)$ given by Eq. (4). Hence we can use the regular sine-Gordon scaling equations [the first two in Eqs. (6)] on the invariant line $x^2(b) - y^2(b) = (h/h_1)^2$, for $b < b_0$. The RG flow is found to be

$$x(b) = \frac{h}{h_1} \coth\left(\frac{h}{h_1}\ln b\right), \quad y(b) = \frac{h/h_1}{\sinh[(h/h_1)\ln b]}.$$
$$(15)$$

Using the Callan-Symmanzik equation for the spin-spin correlations, we have

$$\langle S_+(R)S_-(0)\rangle \simeq \cos[2k_F(h)aR]$$
$$\times \exp\left\{-\int_0^R \frac{db}{b}\left[1 - \frac{1}{2}x(b)\right]\right\} \text{ const,}$$

$$\langle [S_z(R) - M_z][S_z(0) - M_z]\rangle$$
$$\simeq \cos[2k_F(h)aR]\exp\left\{-\int_0^R \frac{db}{b}\left[1 - y(b)\right.\right.$$
$$\left.\left.+ \frac{1}{2}x(b)\right]\right\} \text{ const.}$$
$$(16)$$

For details of the derivation, see Ref. 28. Substitute the RG flow, Eqs. (15), into the above expressions, and we get

$$\langle S_+(R)S_-(0)\rangle_{R < b_0}$$
$$\simeq \frac{\cos[2k_F(h)aR]}{R}\left[\sinh\left(\frac{h}{h_1}\ln R\right)\bigg/\left(\frac{h}{h_1}\right)\right]^{1/2} \text{const,}$$

$$\langle [S_z(R) - M_z][S_z(0) - M_z]\rangle_{R < b_0}$$
$$\simeq \frac{\cos[2k_F(h)aR]}{R}\frac{\tanh[(h/2h_1)\ln R]/(h/2h_1)}{[\sinh[(h/h_1)\ln R]/(h/h_1)]^{1/2}} \text{const.}$$
$$(17)$$

When the magnetic field $h$ vanishes, the anisotropy between the transversal and longitudinal correlations disappears, reproducing the critical sine-Gordon correlation $(-1)^R\sqrt{\ln R}/R$. The result can also be obtained directly by calculating Eqs. (16) on the separatrix line $x^2(b) - y^2(b) = 0$.

Finally we want to make some remarks for the cases when the temperature is finite. Any practical measurement of the magnetic responses can only be done at finite temperatures. So far we have discussed only the field dependences of the physical quantities at $T = 0$. The results are still correct as long as the temperature is low enough such that $h_0/h < T_0/T$ where $T_0$ is a constant. On the other hand, in the weak-field limit $T_0/T < h_0/h$, the temperature plays the role to cutoff. We need then to replace $b_0 \sim h_0/h$ everywhere in the scaling we discussed above by $T_0/T$. The replacement results, in the limit $h \rightarrow 0$, $\pi^2 J\chi_z(T)/(g\mu_B)^2 = 1 + 1/[2\ln(T_0/T)]$ which was found in Ref. 14. The RG flow in the weak-field limit is not critical. Only at length scales shorter than the correlation length determined by the inverse temperature does one observe critical-like behavior, while in a finite magnetic field the flow actually results in a new fixed point with an infinite correlation length.

One possible application of the results presented in the paper is that the asymptotic expression of the susceptibility, Eq. (12) in the weak-field region, can be used to measure the exchange energy $J$ of 1D Heisenberg antiferromagnets. The system can be kept at a low but finite temperature which is above the 3D ordering temperature. When an increasing magnetic field is applied, we will first see the region dominated by the temperature cutoff where $\chi(h)$ changes slowly. After the field becomes strong enough to play the role of cutoff such that $h_0/h < T_0/T$, we are in the regime where the asymptotic expression of $\chi(h)$ describes the magnetic response as long as $h$ is small. Since $\chi(h)$ will increase much faster than that in the first region, we can easily tell the crossover between the two regimes. For an exchange energy $J \sim 2000$ K, a field as strong as $H = 10$ T corresponds to a dimensionless field $h \sim 7 \times 10^{-3}$, which is deep enough in the asymptotic region such that our expression of $\chi(h)$ is a good approximation.

In conclusion, we analyzed the response of an AF Heisenberg chain in external magnetic field. We got the perturbative RG flow equations. The RG flow was found to start near the critical line of the $h = 0$ case and later, at the length scale $h_0/h$, cross over to a region with the fixed point described by a free boson system. We computed the uniform susceptibility $\chi_z(h)$ and compared it with the numerical result. The infinite slope of $\chi_z(h)$ as the field goes to zero was found to originate from a logarithm contribution: $\chi_z(h) \propto \{1 + 1/[2\ln(h_0/h)]\}$. We calculated the spin-spin correlations. Our results are compared with those at finite temperatures in the limit of zero magnetic field. We suggested to determine the exchange energy $J$ by the measurement of the susceptibility $\chi_z(h)$.

*Note added.* After our paper was submitted, we learned about a paper[36] by Affleck and Oshikawa which addresses similar problems, in particular the susceptibility of the Heisenberg model in an external magnetic field. Although the method used by us is different from that by the authors, the results are similar. The main difference is in the proof of the existence of the set of fixed points by us.

The authors would like to thank Professor Joseph L. Birman and Professor B. Sakita for interesting discussions and for reading the manuscript. P.S. wishes to thank Professor V. P. Nair and Professor Stu. Samuel for many useful conversations.

[1] E. Dagotto and T.M. Rice, Science **271**, 618 (1996).

[2] F.D.M. Haldane, Phys. Lett. **93A**, 464 (1983).

[3] K. Hida, J. Phys. Soc. Jpn. **63**, 2359 (1994).

[4] G. Chaboussant, P.A. Crowell, L.P. Levy, O. Piovesana, A. Madouri, and D. Mailly, Phys. Rev. B **55**, 3046 (1997).

[5] C.A. Hayward, D. Poilblanc, and L.P. Levy, Phys. Rev. B **54**, R12 649 (1996).

[6] Z. Weizhong, R.R.P. Singh, and J. Oitmaa, Phys. Rev. B **55**, 8052 (1997).

[7] D.C. Cabra, A. Honecker, and P. Pujol, Phys. Rev. Lett. **79**, 5126 (1997); Phys. Rev. B **58**, 6241 (1998).

[8] T. Sakai and M. Takahashi, Phys. Rev. B **57**, R3201 (1998).

[9] M. Usami and S. Suga, Phys. Rev. B **58**, 14 401 (1998).

[10] K. Tandon, S. Lal, S.K. Pati, S. Ramasesha, and D. Sen, Phys. Rev. B **59**, 396 (1999).

[11] M. Oshikawa, M. Yamanaka, and I. Affleck, Phys. Rev. Lett. **78**, 1984 (1997).

[12] I. Affleck, Phys. Rev. B **41**, 6697 (1990).

[13] S. Sachdev, Phys. Rev. B **50**, 13 006 (1994); S. Sachdev, T. Senthil, and R. Shankar, *ibid.* **50**, 258 (1994).

[14] S. Eggert, I. Affleck, and M. Takahashi, Phys. Rev. Lett. **73**, 332 (1994).

[15] J. Sagi and I. Affleck, Phys. Rev. B **53**, 9188 (1996).

[16] J. Kishine and H. Fukuyama, J. Phys. Soc. Jpn. **66**, 26 (1997).

[17] R. Chrita and T. Giamarchi, Phys. Rev. B **55**, 5816 (1997).

[18] K. Totsuka, Phys. Rev. B **57**, 3454 (1998).

[19] For reviews, see J. Voit, Rep. Prog. Phys. **58**, 977 (1995); V.J. Emery, in *Highly Conducting One-Dimensional Solids*, edited by J.T. Devreese, R.P. Evrard, and V.E. Van Doren (Plenum, New York, 1979), p. 247; J. Solyom, Adv. Phys. **28**, 201 (1979).

[20] H.J. Schultz, Phys. Rev. B **34**, 6372 (1986).

[21] D.G. Shelton, A.A. Nersesyan, and A.M. Tsvelik, Phys. Rev. B **53**, 8521 (1996).

[22] D. Schmeltzer and P. Sun, J. Phys.: Condens. Matter **10**, 4435 (1998).

[23] D.C. Mattis and E.H. Lieb, J. Math. Phys. **6**, 304 (1965).

[24] H. Bethe, Z. Phys. **71**, 205 (1931).

[25] R.B. Griffiths, Phys. Rev. **133**, A768 (1964). The exchange used in this paper was 2$J$ instead of $J$. So when the related results are cited, we rescale them by a factor of **2**.

[26] C.N. Yang and C.P. Yang, Phys. Rev. **150**, 327 (1966).

[27] I. Affleck, D. Gepner, H.J. Schultz, and T. Ziman, J. Phys. A **22**, 511 (1989).

[28] R.R.P. Singh, M.E. Fisher, and R. Shankar, Phys. Rev. B **39**, 2562 (1989).

[29] M. Takahashi, Phys. Rev. B **43**, 5788 (1990).

[30] N.M. Bogoliubov, A.G. Izergin, and V.E. Korepin, Nucl. Phys. B **275**, 687 (1996); V.E. Korepin, N.M. Bogoliubov, and A.G. Izergin, *Quantum Inverse Scattering Method and Correlation Functions* (Cambridge University Press, Cambridge, England, 1996).

[31] A. Fledderjohann, C. Gerhardt, K.H. Mutter, A. Schmitt, and M. Karbach, Phys. Rev. B **54**, 7168 (1996).

[32] P.R. Hammar, M.B. Stone, D.H. Reich, C. Broholm, P.J. Gibson, M.M. Turnbull, C.P. Landee, and M. Oshikawa, Phys. Rev. B **59**, 1008 (1999).

[33] M.C. Cross and D.S. Fisher, Phys. Rev. B **19**, 402 (1979).

[34] F.D.M. Haldane, J. Phys. C **14**, 2585 (1981).

[35] J.M. Kosterlitz, J. Phys. C **7**, 1046 (1974).

[36] I. Affleck and M. Oshikawa, Phys. Rev. B **60**, 1038 (1999).